# SLA ESTABLISHMENT WITH GUARANTEED QOS IN THE INTERDOMAIN NETWORK: A STOCK MODEL [*]

Dominique Barth[1], Boubkeur Boudaoud[1] and Thierry Mautor[1]

[1] PRiSM, University of Versailles-Saint-Quentin, 45 Av. des Etats-Unis  78000  France
```
dominique.barth@prism.uvsq.fr
boubkeur.boudaoud@prism.uvsq.fr
  thierry.mautor@prism.uvsq.fr
```

## ABSTRACT

*The new model that we present in this paper is introduced in the context of guaranteed QoS and resources management in the inter-domain routing framework. This model, called the stock model, is based on a reverse cascade approach and is applied in a distributed context. So transit providers have to learn the right capacities to buy and to stock and, therefore learning theory is applied through an iterative process. We show that transit providers manage to learn how to strategically choose their capacities on each route in order to maximize their benefits, despite the very incomplete information. Finally, we provide and analyse some simulation results given by the application of the model in a simple case where the model quickly converges to a stable state.*

## 1. INTRODUCTION

In the inter-domain network routing, the service Level Agreements (SLAs) correspond to the contracts between Internet Service Providers (ISPs) and their users for services that ISPs provide to their customers. For the Benefit of the ISPs and their users in the inter-domain, the concept of SLA has been introduced. These SLAs are characterized by the informations of guaranteed QoS and accounting of services. They allow to identify and define customer needs. Both QoS-guaranteed services and the accounting cost are critical to the SLA as far as Internet services are concerned [10]. A service contract or SLA in the inter-domain, is a bipartite agreement between customer and service provider. According to [4] the negotiation between both parties (customer and provider) reflects SLS (Service Level Specification) which contains a set of technical parameters that define the level of service provided by a domain. Provisioning is one of the phases that SLA contains in its life cycle [3] and [4], this phase consists in reserving resources and activating the service. In this paper we show how to implement the management of resources and inter-domain QoS guaranteed services, while taking into account the economic constraints of the various operators in the inter-domain network. For this implementation, we have used a new model that we have called **stock model**. In order to reduce the complexity of QoS indicators, we express them by only two parameters corresponding to the equivalent bandwidth and transmission delay of traffic.

In our work, we consider that the customers buy the routes with guaranteed QoS from their transit providers. The goal of purchasing a route by an operator is to allow its customers to reach a desired service in the destination of this route. The purchase of a route is based on the reservation of the guaranteed temporal capacity on this route, the transmission delay of traffic

[*] This work is supported by the FP7 european project "ETICS: Economics and Technologies for Inter-Carrier Services"

DOI : 10.5121/ijcnc.2011.3413

                                                                                                    188

must be also guaranteed on this route. This temporal capacity is available over a precise period of time (eg, a unit of time is one hour), and the transmission delay on this route is defined as the sum of transmission delays between each pair of adjacent nodes from the customer until the destination. For purchasing a route to a given destination, we consider that each domain in the network negotiates with its neighbor (transit provider) the capacity and its availability on this route. In this case, the customer asks to its provider for a **minimum capacity**[1] and a **maximum capacity**[2], it also asks for a **period of availability**[3] of the capacity asked on this route. This customer must also choose a **start time**[4] of a period asked. Each contract established between one customer and its provider for a route to a given destination has the following characteristics: the capacity determined by the provider between the minimum capacity and the maximum capacity, the period of availability and its start time chosen by the customer, the transmission delay of traffic and the price of this route. [2][5][7] and [8] have addressed the comparison with usage dependent pricing. After the establishment of this contract, the customer may negotiate with its transit provider the decrease or the increase of the purchased capacity. In order to keep the same contract with just making a change on the quantity of capacity, this negotiation of capacity should remain in the interval between the minimum capacity and the maximum capacity initially asked. The main raison for which we have based on this interval of capacity for the purchase of a route, is to allow the customer adjustments of capacity on this route to a desired service. In a stock model, each route with its source and its destination, corresponds to a set of contracts between each pair of neighboring operators (customer and provider) from source to destination. An operator that has a route to a given destination must have one single provider, but it can have several customers or no customer. The application of this new model is based on a **reverse cascade** approach: each domain in the network negotiates with its neighbor (provider) the capacity and its availability on the route to a given destination, for its own traffic and the traffic in transit from its neighbors. The negotiations between each two adjacent domains (provider and customer) are based on the exchange of communication messages between these domains. Each communication between two domains (provider and customer) is initiated by the provider that wants to sell capacity on the route with a given destination. Thus this communication is characterized by: the offer of this route by the provider, the demand of capacity on this route by the customer, the confirmation of sale of capacity by the provider and finally the confirmation of purchase of this capacity by the customer. Therefore, in any case if an operator is not a destination (service provider), it cannot propose a route with a guarantee of QoS if it did not buy previously sufficient capacity on this route.

In this model, we apply both learning theory and distributed algorithmic. The algorithms proposed in the model are used on the various communication steps between the nodes in the inter-domain network, and they allow us to determine the characteristics of a route that each node wants to sell or to buy. In this study, after the application of this adequate model proposed that takes the different dimensions of the problem on some specific scenario, primarily by considering the case of one destination and several sources, we focus on its efficiency in the inter-domain network and the satisfaction rate of different operators. we focus also on the analysis of the stabilizing behavior of the distributed system and the convergence time in the case where a model converges to the stable states.

Some authors [1][13] have considered a cascading connectivity model in the inter-domain networks. In [1], both a theoretical game and a distributed algorithmic approach are presented for the transit price negotiation problem in the inter-domain routing framework and the authors have highlighted situations where cooperation is possible in order to maintain higher prices. In this paper, we consider as in [1] that the traffic dedicated to a given destination can be routed only through a single provider: only one path is chosen to each destination. In [13], a cascaded connectivity model in the inter-domain network is introduced and analyzed for the establishment of SLAs using game theory with realistic rules. This work has shed some light over inter-domain routing, the authors have showed that there is a link between SLA cascading and the business models of carriers. The difference between the analysis in this work and our proposition is that we use a reverse cascade model. Another difference is that in [13], SLAs are established at the end of all negotiations between all domains, while in our case, SLAs are established after each negotiation between provider and customers. The impact of this reverse cascading model on the structure of the Internet has never been analyzed in detail to our knowledge.

---

[1] minimum capacity asked by an operator to its neighbor on a route to a given destination
[2] maximum capacity asked by an operator to its neighbor on a route to a given destination.
[3] A period asked by an operator to its neighbor on a route to a given destination.
[4] beginning of the period of availability of a route asked by a customer to its neighbor.

Authors have proposed and used a decentralized learning algorithm in [9]. Considering games with incomplete information, this algorithm is introduced in order to reach a Nash Equilibrium. Authors have given a proof to ensure that convergence can take place into a Nash equilibrium when players follow the learning rule. Similarly to the algorithm taken in [9], an approach has also been considered in [6] for the game where players select a number of parallel TCP sessions to open. Using game theory and assuming that demand of each service is determined by prices, the researchers in [11][12] have studied the revenue management competition in inventory and pricing without considering demand overflow. Learning theory is also used In [14] to study the game related to distributed discrete power control in wireless networks. Authors have proved that the proposed learning algorithms converge only to a point which is a Nash equilibrium in [9] and [14].

Structure of the paper: Section (2) is dedicated to the description of the model. Before describing the stock model and the reverse cascade approach in this section, we first describe the inter-domain network and we explain what is a contract between two domains for the purchase of route. In section (3), we present the process of different communication steps between neighbors for the purchase of route. After, we present in section (4) the policies applied by each node in the model. Finally we present in section (5) the simulations analysis under one only topology and finish by the conclusion in section (6).

## 2. MODEL DESCRIPTION

### 2.1. The network model

We begin this section by modeling the inter-domain network, we model the network by an undirected graph $G(V,E)$, where sets $V$ and $E$ respectively denote the ASs[5] and inter-domain links. We consider that the vertices of the graph are weighted by their capacity given by the function $cap: V \rightarrow N^*$, and their delay given by $d: V \rightarrow N^*$. These functions represent respectively the maximum quantity of traffic that can pass through node with a guaranteed QoS and the time required for sending traffic from node to one of its neighbors. We can separate the capacity used by each node $v$ into three parts (see Fig1):

---

[5] Autonomous System. We assume in the following of this document that each domain in the network has only one single AS. Thus an AS, a domain and an operator correspond all to a node

- $cap\_IN(v)$: the capacity of traffic sent by other nodes with $v$ as destination.
- $cap\_OUT(v)$: the capacity used by node $v$ to send its own traffic to other destinations.
- $cap\_IN\_OUT(v)$: the capacity used by node $v$ to route some traffic from some of its neighbors to other ones.

We note that.

$$cap\_IN + cap\_OUT + cap\_IN\_OUT(v) \leq cap(v)$$

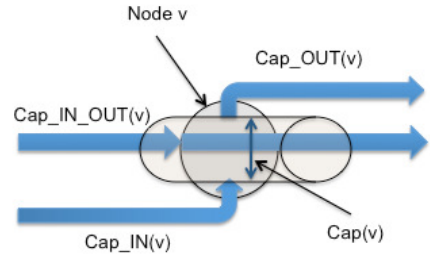

Fig1. The different categories of capacity used by node $v$

We assign to each vertex $v$ of a graph $G$, a number $num(v)$ from 1 to $n$ and we denote by $\Gamma_{G(v)}$ the set of neighbors of $v$. We focus on the traffic flow between the nodes of the graph that requires a guarantee of QoS. A variable $S$ represents the stage period on which inputs (graph, traffic Matrix) are stable, In such case the traffic will be denoted as $M^S[v,u]$ corresponding to the traffic to be sent from $v$ to reach a service of node $u$ (service provider) at stage $S$. The capacity that node $v$ wishes to buy at stage $S$ for its own traffic is easy to compute and is given by the expression $M^S(v) = \sum_{u \in V} M^S[v,u]$ ($M^S(v) \leq cap(v)$). However, the necessary capacity for the traffic coming from other operators and passing through node $v$ is not directly known by this node. We also consider that for each node $v$, we have a vector of margins $marg_v$ of dimension $n$, where each coordinate $marg_v(u)$ of this vector has a discrete value and corresponds to the local price per unit charged by node $v$ for each unit of capacity sold on the route with destination $u$. Again, if this value could vary, it will be denoted as $marg_v^S(u)$ in the vector $marg_v^S$ at stage $S$. We consider that the unit price to be paid by the node $v$ to its provider for buying route with destination $u$ is equal to the sum of margins of all nodes on this route from $v$ to $u$.

Finally, we associate to each node $v$ a constant $utility(v)$, which represents the utility per unit of the capacity $M^S[v,u]$ purchased by this node $v$ at stage $S$ to any destination $u$. We consider that if the cost of route at stage $S$ is upper than or equal to $utility(v)$, $v$ will refuse to send its own traffic. And we also associate to each node $v$ a constant $\max\_delay(v)$, which represents the maximum time required to transmit packets from the source $v$ to any destination $u$. We consider that if the delay of the route at each stage $S$ is upper than $\max\_delay(v)$, node $v$ will refuse to send its own traffic at this stage.

During the stage $S$, the inputs of the problems are stable, and the game can model interaction between nodes during the negotiation of capacity on the route with a given destination.

## 2.2. Contract Definition

Each SLA between two neighboring ASs corresponds to the bilateral contract for purchasing of route to the given destination, this route is purchased by the customer from its provider. This contract defines the commitment of each AS to another one, between the two ASs (the provider that sells and the customer that buys a route to the given destination). Thus each contract related to each route purchased by the customer from its provider has the following characteristics: the capacity, the delay, the cost, the availability, the start time and the provider of this route. We denote the contract established between a customer $v$ and its provider for a route with $u$ as destination by "$Contract_{prov_{v,u},v}(u)$"; $prov_{v,u}$ is the provider of a route purchased by $v$, i.e a route purchased by $v$ is charged by the provider $prov_{v,u}$. This contract with its characteristics are denoted "$Contract_{prov_{v,u},v}(u)=\{poss_{v,u}, delay_{v,u}, \cos t_{v,u}, \#blok_{v,u}, start_{v,u}, prov_{v,u}\}$". This contract indicates that the customer $v$ has bought to its provider $prov_{v,u}$ a route with $u$ as destination. The characteristics of this route are: the capacity $poss_{v,u}$, the transmission delay from $v$ to $u$ $delay_{v,u}$, the unit price $\cos t_{v,u}$, the availability period $\#blok_{v,u}$ and the beginning time $start_{v,u}$.

### 2.3. Stock Model

In this model, each AS may sell to its neighbors a part or all the capacity on each of its routes. As an AS may sell the route of its own services if it is destination (service provider), it may also sell the routes that it has bought to other destinations. An AS cannot sell the route to its neighbor, which is its own provider of this route. We have already seen that if an AS is not a destination (service provider), it cannot propose a route with a guarantee of QoS if it did not buy previously sufficient capacity on this route. Fig2 shows on more detail **how a customer $w$ can buy the capacity from its neighbor $v$ on the route with $u$ as destination.**

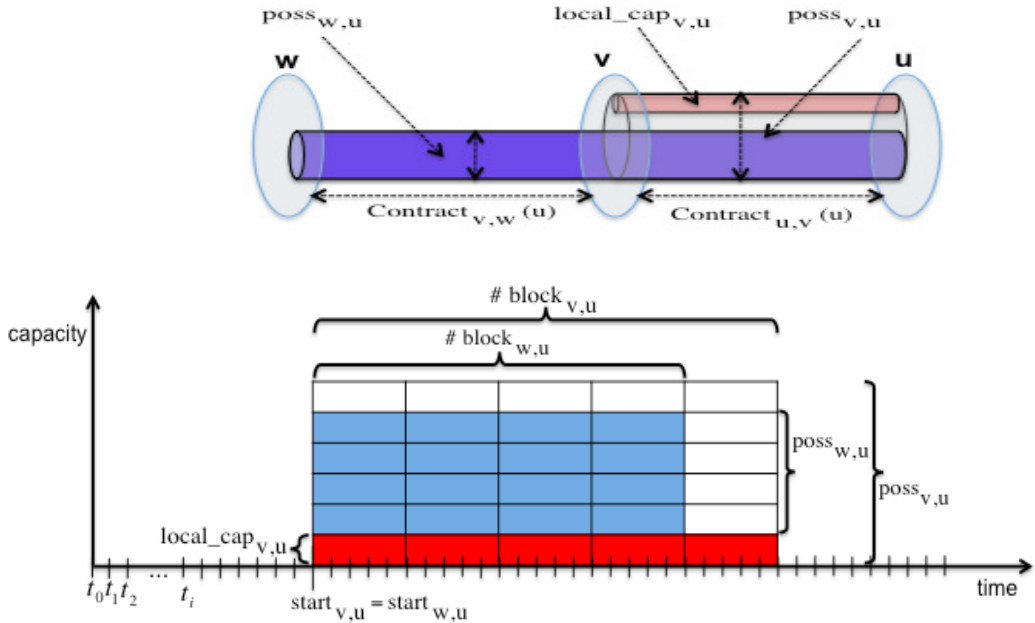

Fig2. The different categories of capacity used by node $v$

In the example (Fig2) we assume that before a time $t_i$, the node $v$ has established with $u$ a contract $Contract_{u,v}(u)$. This contract indicates that node $v$ has bought a route with guaranteed QoS from its service provider $u$, so the destination of this route is $u$. We assume that the capacity on this route is $poss_{v,u}$. This capacity $poss_{v,u}$ is available during the period $\#blok_{v,u}$ that begins at time $start_{v,u}$. After the purchase of this route, the node $v$ decides to keep a part of the capacity purchased to its own traffic and to sell the rest (the unused capacity bought or free capacity). We denote this part of capacity kept to its own traffic $local\_cap_{v,u}$. Always in the case of this example, to sell the unused capacity that $v$ has bought, it proposes this capacity to its neighbor $w$. We assume that the offer is interesting for the node $w$, so this node $w$ buys a capacity $poss_{w,u}$. Then both nodes $v$ and $w$ establish a contract $Contract_{v,w}(u)$ for the route with guaranteed QoS that $w$ has bought from $v$.

Now, we show in the following example (Fig3) **how an AS can reserve capacity on a route to destination $u$ at stage $S$. This capacity is reserved for itself and also for more than one of its neighbors.** i.e after purchasing capacity by a node $v$ from its provider $prov_{v,u}^S$, the node $v$ keeps a part $local\_cap_{v,u}^S$ to its own traffic, then it resells to its own neighbors the maximum that it can of remaining capacity on this route. In this example, $v$ resells for each one of its neighbors $x$, $y$ and $z$ respectively the capacities $poss_{x,u}^S$, $poss_{y,u}^S$ and $poss_{z,u}^S$. These capacities are available respectively during the periods $\#blok_{x,u}^S$, $\#blok_{y,u}^S$ and $\#blok_{z,u}^S$. Each of these last ones has its beginning time, respectively $start_{x,u}^S$, $start_{y,u}^S$ and $start_{z,u}^S$. The white portion in the figure is the available capacity during one block of time that node $v$ can not resell, because none of its neighbors has asked for this capacity.

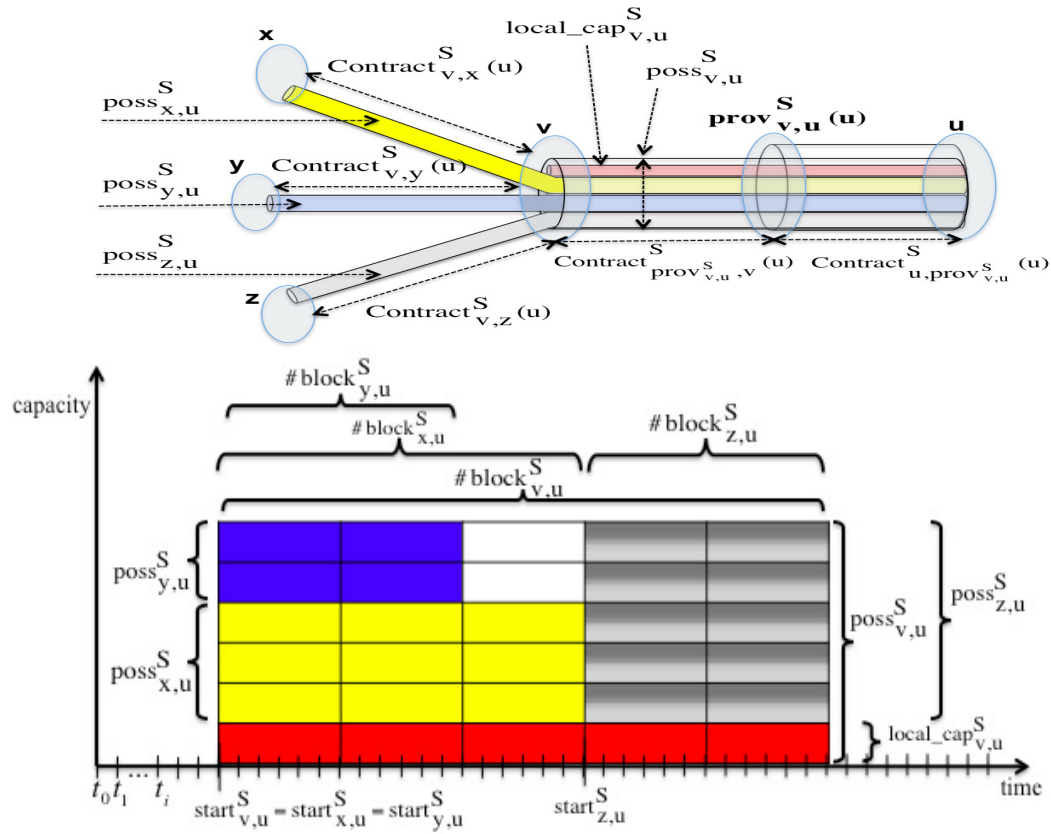

Fig3. The capacity and its availability purchased by each node $v$, $x$, $y$ and $z$ on the route with $u$ as destination at stage $S$

Considering just the nodes $x$ and $v$ in (Fig3), we show **the information exchanged** at each time during the stage $S$ between these nodes before the establishment of contract $Contract_{v,x}^S(u)$. We assume that node $v$ has established $Contract_{prov_{v,u},v}^S(u)$ before a time $t_i$. Firstly at time $t_i$, the node $v$ sends to its neighbors a message in the form of an offer of available capacity on its route with $u$ as destination. Secondly at time $t_{i+1}$, the node $x$ sends to $v$ a message asking for capacity between minimal capacity $cap\_min_{x,u}^S$ and maximal capacity $cap\_max_{x,u}^S$. In this message, node $x$ must choose the availability period $\#blok_{x,u}^S$ and the start time $start_{x,u}^S$ of this capacity asked. Thirdly at time $t_{i+2}$, node $v$ sends a positive or a negative response to $x$. In the case of this example, it sends a positive response and it must define the capacity that it can sell to this neighbor $x$. The capacity according to $x$ is $poss_{x,u}^S$ and should be given between $cap\_min_{x,u}^S$ and $cap\_max_{x,u}^S$. And finally at time $t_{i+3}$, node $x$ sends a confirmation response to $v$, this response may be positive or negative. In this example, $x$ sends to $v$ a positive response and both nodes establish a contract $Contract_{v,x}^S(u)$.

After the establishment of this contract, node $x$ can negotiate with its transit provider $v$ the decrease or the increase of the capacity purchased. In order to keep the same contract with just making a change on the quantity of capacity, this negotiation of capacity should be in the interval between the minimum capacity and the maximum capacity. i.e. if the customer wants just change the quantity of capacity purchased on this route, it cannot asks to its provider for more than the maximum capacity or for unless than the minimum capacity already asked. In the case where node $x$ negotiates to reduce its capacity, two cases may occur: if the negotiation is done before start time $start_{x,u}^S$, node $x$ can reduce its capacity without paying the capacity dropped on this route. this capacity dropped is the capacity unused by $x$, so node $x$ must give back this capacity unused to $v$. If the negotiation is done after start time $start_{x,u}^S$ and during the availability period $\#blok_{x,u}^S$, node $x$ can reduce its capacity but it must pay a penalty that its provider requires. To understand this exchange of information the general case, we will detail it in the section: Communication steps between neighbors in the model.

Now, we show **the process of reservation of capacity** by the nodes on the routes to a given destination using a **reverse cascade approach**. During the stage $S$, the inputs of the problem are stable and each As that wants sell capacity on its route it announces this route to its neighbors with the corresponding characteristics. When an As decides to buy a route with guaranteed QoS from its neighbor, it can itself announces its route to its own neighbors with the new characteristics. In (Fig4), the destination node $u$ offers a route to its neighbors and then selects a set of customers $(v, y, s)$, these customers do the same with their own customers ($v$ offers to $w$, $y$ offers to $w, z$ and $s$ offers to $z$), and so on until the moment when there is not the offer of capacity on the routes towards the destination $u$.

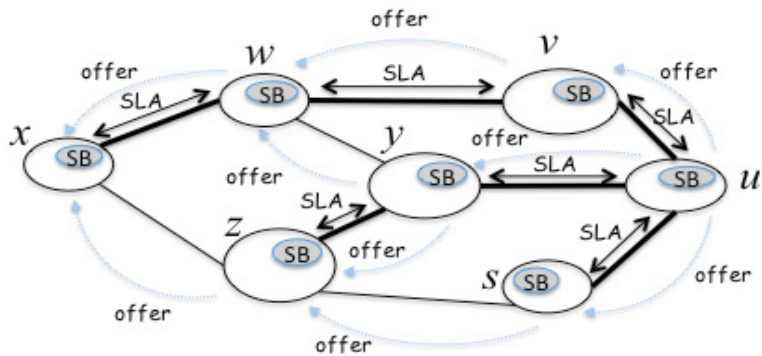

Fig4. Reservation of capacity by the operators on the routes to destination $u$ using a reverse cascade approach

We have already seen that each SLA corresponds to the bilateral contract between the customer and its provider for purchasing route to the given destination (the destination is $u$ in this example). These SLAs are established just after the several steps of communication between each two neighboring operators. Considering just the nodes $x$, $w$, $v$ and $u$ in (Fig4), the figure (Fig5) shows the running order of offer queries, demand queries and confirmation queries between these operators for the acquisition of capacity. For example, before that the node $v$ establishes contract with the service provider $u$, it cannot offer capacity to its neighbot $w$ on the route with $u$ as destination. A contract between two domains for the purchase of route is always preceded by negotiation between these nodes, this negociation is based on the communication messages that we have mentioned earlier.

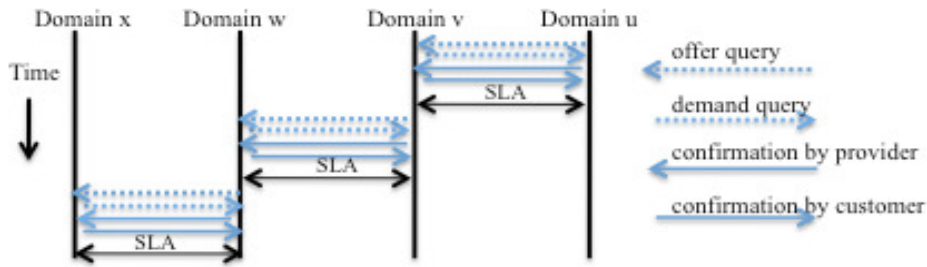

Fig5. The order of queries in time

In this model, the domain that buys route to the given destination will have either positive or negative benefit. Among the causes of negative benefits we can cite: 1) The domain that has bought too much capacity and it has not managed to resell this capacity to its neighbors. 2) The domain that has not bought enough capacity in order to satisfy the demands of its neighbors. These consequences are due to competition between domains, but also in the fact that in this model each transit domain buys capacity in blind. I.e. each domain that wants to purchase capacity on the route to the given destination, this domain does not know the needs of its neighbors for this route. Therefore, the introduction of iterative procedures for learning in this model is justified and motivated.

Thus, using this iterative procedures, the problem that faces each domain consists: in the case where this domain is in competition with other domains for purchasing route to the given destination, how to fix the interval of capacity and its period of availability in order to be chosen for purchasing this route. In the case where this domain is in competition with other domains for selling route to the given destination, how to fix its transit price and how much capacity to assign to each neighbor in order to be chosen to route the traffic of the greatest possible number of neighbors that maximize its profit.

In this model, we have seen that the negotiation follows a reverse cascade approach from the destination backward to the source, where each domain in the path plays both the customer and the provider role. The objective of each domain at each new stage is clearly to maximize its own benefit with using the informations of the previous stage. In order that each domain maximizes its profit corresponding to the given route, it propose attractive capacities with attractive transit prices, but also by choosing itself the transit providers that offer the routes with a lot of capacities and with the lowest prices. Therefore, each domain must evolve some parameters on this route (price, capacity, delay, availability). This model converges to the **stable state** when each domain in the network has not interest to deviate from its chosen strategy, so the stable state corresponds to **NE (Nash Equilibrium)**. In the cases where this model converges to the stable state, the satisfaction rate is interesting if an important number of domains in the network are satisfied.

## 3. COMMUNICATION STEPS BETWEEN NEIGHBORS IN THE MODEL

The exchange of information between the operators follows an asynchronous communication model with discrete events and we distinguish four main events according to the role of the (provider or customer), as shown in Fig6. These events are controlled by the operator policies that will be defined in Model policies section.

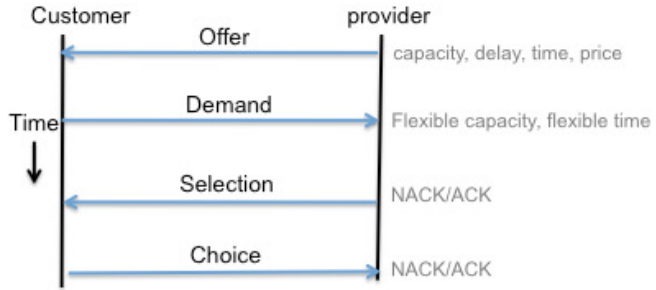

Fig6. Communications between provider and customer

### 3.1. The offer step

The providers announce the characteristics of their own services if they are destinations (services providers), else they announce the characteristics of routes they have bought (capacity, delay, price, destination and availability) to their neighbors which are not their own providers. The offer of capacity on the route to destination $u$ is always initiated by the destination node $u$ (service provider), and this node must have at least one service. The provider node can initiate this event when it wants sell a capacity on its route.

At each stage $S$ and for each destination $u$, the sextuplet "$Contract_{prov_{v,u}^S,v}(u) = \{poss_{v,u}^S, delay_{v,u}^S, \cos t_{v,u}^S, \#blok_{v,u}^S, start_{v,u}^S, prov_{v,u}^S\}$" indicates that the node $v$ has a route to $u$ with capacity $poss_{v,u}^S$ and transmission delay $delay_{v,u}^S$. This route was sold by the provider $prov_{v,u}^S \in \Gamma_{G(v)}$ for a price $\cos t_{v,u}^S$ and the number of blocks of time $\#blok_{v,u}^S$ when this route is available. The availability of this route begins at time $start_{v,u}^S$. We consider that the following properties have to be verified:

$$\sum_{w \in Cust_{v,u}^S} poss_{w,u}^S \leq poss_{v,u}^S, \text{ where } Cust_{v,u}^S = \{w \in \Gamma(v) | prov_{w,u}^S = v\}$$ is the set of customers of the node $v$ at step $S$ for the route to destination $u$.

$$\sum_{u \in V} poss_{v,u}^S \leq cap(v)$$

$$\#blok_{v,u}^S \leq \#blok_{prov_{v,u}^S,u}^S$$

$$start_{v,u}^S \geq start_{prov_{v,u}^S,u}^S$$

$$\#blok_{v,u}^S + start_{v,u}^S \leq \#blok_{prov_{v,u}^S,u}^S + start_{prov_{v,u}^S,u}^S$$

Initially after the node $v$ has purchased a route with destination $u$ at stage $S$, it keeps a certain capacity for its own traffic if it is profitable and sells to its neighbors what it can of the

196

remainning capacity on this route it has bought. We consider that the following properties have to be verified:

- $local\_cap_{v,u}^S = \begin{cases} M^S[v,u] \text{ if } \cos t_{v,u}^S < utility(v) \\ 0 \text{ if } \cos t_{v,u}^S \geq utility(v) \end{cases}$

- $local\_cap_{v,u}^S \leq (poss_{v,u}^S - \sum_{w \in Cust_{v,u}^S} poss_{w,u}^S)$

Let us remark that at each stage $S$, the available capacity on the route with $u$ as destination is:
$free\_cap_{v,u}^S \leq poss_{v,u}^S - (local\_cap_{v,u}^S + \sum_{w \in Cust_{v,u}^S} poss_{w,u}^S)$

We suppose that a node $v$ has a route to the destination $u$ at step $S$, $rout_{v,u}^S = \{poss_{v,u}^S, delay_{v,u}^S, \cos t_{v,u}^S, \#blok_{v,u}^S, start_{v,u}^S, prov_{v,u}^S\}$.
Therefore, the node $v$ sends an offer request "$offer\_route_{v,u}^S$" to each node $w$ such that $w \in \Gamma_{G(v)}$ and $w \neq prov_{v,u}^S$.
$offer\_route = \{free\_cap_{v,u}^S, new\_delay_{v,u}^S, price_{v,u}^S, \#blok_{v,u}^S, star$
such that: $\begin{cases} new\_delay_{v,u}^S = delay_{v,u}^S + d(v) \\ price_{v,u}^S = \cos t_{v,u}^S + m\arg_v^S(u) \end{cases}$

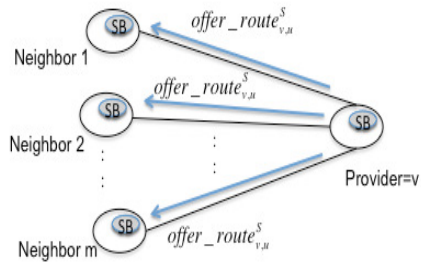

Fig7. The offer step

## 3.2. The demand step

This event corresponds to the sending of requests by the customer $v$ to the potential providers, these requests are related to ask for buying of a route for a given destination $u$ by the customer $v$. At stage $S$, each node $v$ that wants to buy a route to a destination $u$ sends a demand query "$demand\_route_{v,u}^S$"\$ to one of its neighbors in the set $\Omega$, that offered this route (see Fig8).
$demand\_route_{v,u}^S = \{[cap\_\min_{v,u}^S, cap\_\max_{v,u}^S], \#blok_{v,u}^S, start_{v,u}^S\}$.

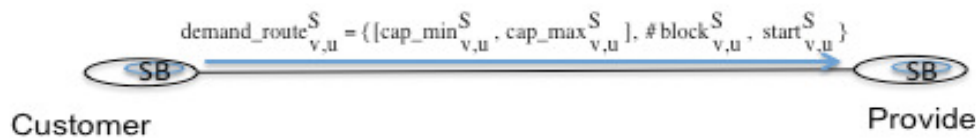

Fig8. The demand step

$\Omega \in \Gamma_{G(v)}$: Set of the neighbors of node $v$ that offer routes with $u$ as destination.

Therefore, in this case, each operator $v$ asks at stage $S$ to its neighbor $p$ in the set $\Omega$ for buying a route with the capacity $poss_{v,u}^{S}$ ($cap\_min_{v,u}^{S} \leq poss_{v,u}^{S} \leq cap\_max_{v,u}^{S}$), for a given period $\#blok_{v,u}^{S}$ and destination $u$, such that ($\#blok_{v,u}^{S} \leq \#blok_{p,u}^{S}$) and ($start_{v,u}^{S} \geq start_{p,u}^{S}$) (See Fig9).

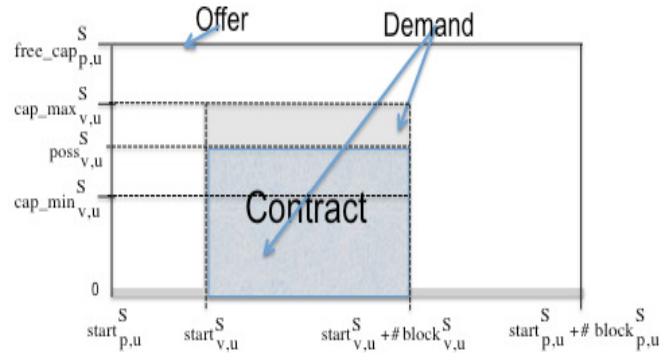

Fig9. The capacity in the interval $[cap\_min_{v,u}^{S}, cap\_max_{v,u}^{S}]$ that the operator $v$ asks to its neighbor $p \in \Omega$ on the route with destination $u$, and for the period $\#blok_{v,u}^{S}$

### 3.3. The selection step

This event is initiated by the provider node $v$ as an answer to the requests of ask for route with destination $u$ in the demand step. This response may be positive or negative. In case of positive answer for a given customer $w$ that has asked to the node $v$ for the route with destination $u$, the node $v$ must determine at stage $S$ the capacity on the route with destination $u$ that can be sold to this customer $w$. Therefore, the provider sends to this customer $w$ a message "$Ok(poss_{w,u}^{S}, \#blok_{w,u}^{S})$", $cap\_min_{w,u}^{S} \leq poss_{w,u}^{S} \leq cap\_max_{w,u}^{S}$. (See Fig10).

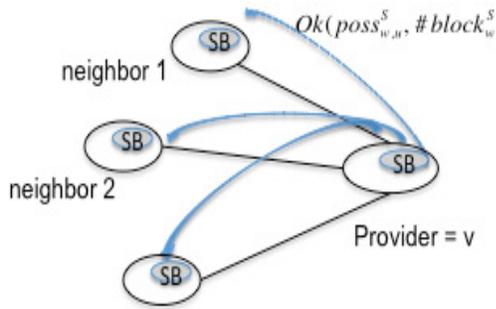

Fig10. The selection step

The provider chooses its neighbors in order to maximize its benefice. Recall that $Cust_{v,u}^{S} = \{w_1, w_2, ..., w_k\}$ the set of customers that the provider $v$ wants to satisfy at stage $S$, the following conditions must be verified:

$$\begin{cases} \sum_{w \in Cust^S_{v,u}} poss^S_{w,u} \leq free\_cap^S_{v,u} \\ cap\_\min^S_{w,u} \leq poss^S_{w,u} \leq cap\_\max^S_{w,u}, \forall w \in Cust^S_{v,u} \\ \#blok^S_{w,u} \leq \#blok^S_{v,u}, \forall w \in Cust^S_{v,u} \\ start^S_{w,u} \geq start^S_{v,u}, \forall w \in Cust^S_{v,u} \\ start^S_{w,u} + \#blok^S_{w,u} \leq start^S_{v,u} + \#blok^S_{v,u}, \forall w \in Cust^S_{v,u} \end{cases}$$

**The selection of the customers by the provider:**

When the provider node $v$ receives demands of route to the destination $u$ by its neighbors in the set $\{w_1, w_2, ..., w_i, ...\}$, it classifies the nodes that ask for this route according to the following criteria:

1. The greatest benefice that the provider $v$ can have about the choice of each node $w_i$.

2. The numbering of nodes $num(w_i)$ in case of equality of benefices, the node $v$ selects its customers in increasing order of their numbers $num(w_i)$.

And afterward, compared to the free capacity in this route of the provider, it satisfies all or a certain number of its neighbors $\{w_1, w_2, ..., w_i, ...\}$ by respecting the classification of these last ones.

Determination of benefice that a customer $w$ can cause to its provider $v$ at stage $S$ (A benefice of the provider $v$ about the capacity that it sold to the customer $w$):

$$Benef^S_{v,u}(w) = (\#blok^S_{w,u}) * (poss^S_{w,u}) * (m\arg^S_{v,u})$$

### 3.4. The choice step

This event is initiated by the customer node as an answer to the positive selection query sent by the provider in the selection step. This response may be positive or negative (see Fig11). Let us note that each node that wishes to change of provider for the route with destination $u$, releases its route only when it receives a confirmation of sale from its new provider.

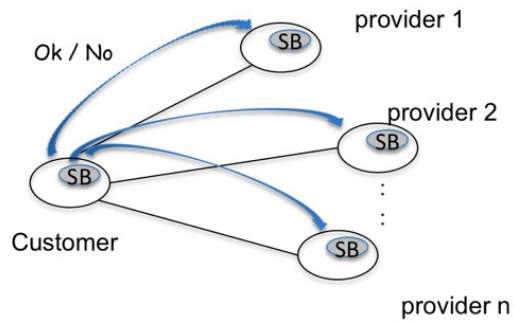

Fig11. The Choice step

**The choice of the provider by the customer:**

When the customer node $v$ is selected for the first time by the provider node $p_1$ offering the route $r_1$ to the destination $u$ at stage $S$ (i.e. it does not already have a route to the destination $u$), it sends the positive response to this provider $p_1$, so $prov^S_{v,u} = p_1$.

And afterward, when the customer is selected during the same stage $S$ by another provider $p_2$ for which it demanded the route $r_2$ with the same destination $u$ (i.e. it does already have the route $r_1$ to the destination $u$ buying by $p_1$, and the availability of this route is not yet begun), we propose three solutions that correspond to the three following models:

a) **The open model:** If the new route $r_2$ is more interesting than its route $r_1$, the customer $v$ releases its route $r_1$ and it sends the positive response to the new provider $p_2$, so $prov^S_{v,u} = p_2$. This model is applied in simulations of our work.

b) **The blocked model:** The customer $v$ that has already chosen its provider $p_1$ cannot release its route $r_1$. So, it never sends another demand of route and so it will not be selected again by another provider.

c) **The model with penalty:** If the new route $r_2$ is more interesting than its route $r_1$. The customer $v$ can release the old route $r_1$ and it sends the positive response to the new provider $p_2$ ($prov^S_{v,u} = p_2$), but the customer $v$ must pay the penalty that the previous provider $p_1$ requires.

## 4. MODEL POLICIES

In this section, we present the different policies applied by the operators in the stock model.

### 4.1. The offer route policy

This policy allows each provider at each stage $S$ to define the quantity of capacity available for the sale on its routes. We can have two cases for the route with $u$ as destination:

1. If the provider node is the destination node, the available capacity that this node offers to its neighbors is equal to its free capacity ($free\_cap^S_{v,v} = poss^S_{v,v}$).

2. If the provider node $v$ is different from the destination node $u$, the available capacity that this provider node $v$ offers to its neighbors is the difference between the capacity purchased on this route and the capacity that it wants to keep for its own customers ($free\_cap^S_{v,u} = poss^S_{v,u} - local\_cap^S_{v,u}$).

### 4.2. The fixation price policy for an offer of route

This policy allows to define the price of each route at each stage $S$. Each provider $v$ that wishes to sell the capacity on its route to destination $u$, it must apply this policy to fix the price of this route. The price that $v$ must fix on this route is equal to the sum of the price that it has bought this route and its margin. So by deduction, the idea in this policy is to fix the margin. Two cases can occur in this policy, before presenting these cases, we show how to determine the benefice of $v$ after the purchase and the sale of capacity on route with $v$ as destination.

$$\text{Benef}^S_{v,u} = [\sum_{w \in \text{Cust}^S_{v,u}} (\#\text{blok}^S_{w,u}) * (poss^S_{w,u}) * (cost^S_{w,u})]$$
$$+ [(\#\text{blok}^S_{v,u}) * (local\_cap^S_{v,u}) * (\text{utility}(v))]$$
$$- [(\#\text{blok}^S_{v,u}) * (poss^S_{v,u}) * (cost^S_{v,u})]$$

**The case where this margin is a constant:** In this case, all nodes choose their margins and these margins do not move during time.

**The case where this margin is not a constant:** In this case, at the first negotiation stage $(S = S_0)$, each node fixes its margin to 1. And afterward, two cases can occur at each stage $(S / S \neq S_0)$:

1. If the provider node $v$ has sold all or a part of its capacity bought at the stage $(S-1)$ and that its benefices were positive ($Benef_{v,u}^{S-1} > 0$: the benefices brought by $v$ for purchasing the route with destination $u$ at stage $(S-1)$), so at this stage $S$ it increases the margin of one unit to hope to have more benefices, $m\arg_v^S(u) = m\arg_v^{S-1}(u) + 1$.

2. If the provider node $v$ has not sold its capacity bought on this route at the stage $(S-1)$ and ($m\arg_v^{S-1}(u) > 1$). We interpret this case by the fact that the competitors of $v$ have proposed cheaper prices on this route at stage $(S-1)$. So at this stage $S$, it decreases the margin of one unit to hope to win against its competitors, $m\arg_v^S(u) = m\arg_v^{S-1}(u) - 1$.

### 4.3. The capacity demand policy

This policy allows each customer at each stage $S$ to define the interval of capacity on each route that it wants to buy. We have already seen that during the phase of communication at each stage $S$ between two nodes (the provider $p$ and the customer $v$) for purchasing of route with destination $u$, the customer demands the capacity for its route on an interval, such that :

$0 \leq cap\_\min_{v,u}^S \leq cap\_\max_{v,u}^S \leq \min(cap(v), free\_cap_{p,u}^S)$, this policy allows the customer $v$ how to determine its $cap\_\min_{v,u}^S$ and $cap\_\max_{v,u}^S$.

**Fixation of $cap\_\min$:** During the first stage $(S = S_0)$, each node $v$ that wants to ask to its neighbor $p$ for the capacity on the route with destination $u$, it fixes its minimal capacity ($cap\_\min_{v,u}^{S_0} = M^{S_0}[v,u]$ if $\cos t_{v,u}^{S_0} < utility(v)$ or $cap\_\min_{v,u}^{S_0} = 0$ if $\cos t_{v,u}^{S_0} > utility(v)$). And afterward, several cases can appear at each negotiation step $(S / S \neq S_0)$.

1. If the node $v$ has managed to sell its capacity bought at stage $(S-1)$ and that its benefices were positive ($Benef_{v,u}^{S-1} > 0$), so at this stage $S$ it requires the same minimal capacity that at stage $(S-1)$. $cap\_\min_{v,u}^S = cap\_\min_{v,u}^{S-1}$.

2. If the node $v$ has not managed to sell its capacity bought on this route at stage $(S-1)$. We interpret this case by the fact that $v$ did not have enough capacity on its route at stage $(S-1)$, i.e. the competitors of $v$ have proposed more capacity on this route or the neighbors of $v$ have required more capacity than what was proposed by $v$. So during this stage $S$, it increases if it can the minimal capacity of one unit to hope to have more capacity and to win against its competitors, $cap\_\min_{v,u}^S = cap\_\min_{v,u}^{S-1} + 1$. If $v$ cannot increase its minimal capacity, i.e it has demanded the maximum that was offered by its provider at $(S-1)$. So during this stage $S$ it asks for $cap\_\min_{v,u}^S = M^S[v,u]$ if $utility(v) > \cos t_{v,u}^S$ or $cap\_\min_{v,u}^S = 0$ if $utility(v) < \cos t_{v,u}^S$ to avoid negative benefice.

**Fixation of $cap\_\max$:** At each stage $S$, each node $v$ asks for the maximum capacity possible that it can. Therefore, it asks for the minimum between the capacity that it can transit and the capacity that has been proposed by its provider.
$cap\_max_{v,u}^{S} = \min(cap^{S}(v), free\_cap_{p,u}^{S}), \forall S$.

### 4.4. The customers' selection policy by the provider

This policy defines the order in which the requests of demand of acquisition of capacity are served by the provider node $v$. This policy is used when the capacity required by the customers exceeds the capacity available on this route. This policy depends on both quantities $cap\_min_{w,u}^{S}$ and $cap\_max_{w,u}^{S}$.

At each stage $S$, each provider $v$ tries to attribute a minimal capacity required by each of its customers $w_i$ in the order $(cap\_min_{w_1} \leq cap\_min_{w_2} \leq ... \leq cap\_min_{w_i} \leq ...)$. And afterward, provider $v$ divides its remaining capacity on the number of customers. Always by respecting the previous same order of the customers, it adds to each of its customers a part of this remaining capacity. This part should not exceed the difference between $cap\_min_{w_i}^{S}$ and $cap\_max_{w_i}^{S}$.

## 5. SIMULATION ANALYSIS

Our objective here is to study the stabilizing behaviour of the distributed system under the capacity adjustment strategy. Simulation is done using C. We consider the topology given in (Fig12), vertices of the graph are weighted by their capacity and their transmission delay.

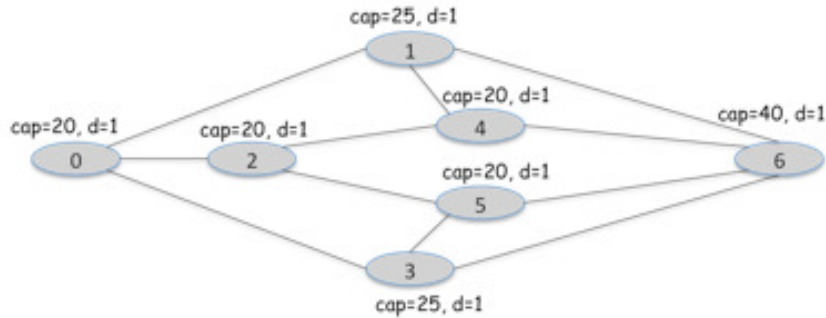

Fig12. The proposed topology

In this example, we consider node 6 as destination and transmission delay of traffic from each node to its neighbors is fixed to 1 unit. We have chosen capacity of each node as shown in the graph in order to have competition between nodes (eg: $cap(1) = cap(3)$ because node 1 and node 3 have a common neighbor node 0, this neighbor will choose the best offer for buying route to node 6). We assume that margin of each node associated to this destination is fixed to 1 unit. The availability period of each route proposed to this destination is fixed to the available capacity chosen by the service provider (destination). So each node that wants to buy route to this destination chooses the same availability period proposed by its neighbor. Therefore, the availability periods ($\#blok$) are the same and the beginning time ($start$) are also the same on all routes to this destination. The utility that each node fixes in order to buy route is 5, and the maximum delay of traffic is 10. Recall that each node that receives an offer of route with the price that exceeds its utility, refuses to buy this route. Also each node that receives an offer of route with transmission delay that exceeds its maximum delay refuses to buy this route. Thus we choose these utilities and these maximum delays for nodes in order that these nodes accept offers of route in the most cases. In the traffic matrix, we focus on the necessary capacity of each source (each node) to transit its own traffic to node 6. We fix these components to 3, i.e $M^S[v,6] = 3, v \in \{0,1,2,3,4,5\}, \forall S$.

The figures (Fig13 and Fig14) show the benefit of each node at each stage. We clearly see in (Fig13) that node 6 increases its benefit from the first stage to the stage 25, this increase of benefit can be interpreted by the fact that all nodes increase their minimum capacities. After the stage 25, the benefit of node 6 is reduced because some nodes cannot obtain their required capacities. ie, nodes that ask for a lot of capacity can not manage to resell all this capacity, and the destination node has not obtained benefit about the capacity given back by these nodes. In (Fig14), the variation in benefits shows that there is a competition between nodes, (eg, competition between nodes 1 and 3, the node that will sell route to node 0 is the node that buys more capacity from node 6. In general, the cause of this change is the change of the minimum capacity asked by each node. At stage 36, all nodes cannot increase their minimum capacities in order to hope for more profits. So at this stage, each node asks for minimum capacity that what it has sold at the previous stage. So at the next stages no node has interest to deviate from its strategy (minimum capacity asked), and the stable state is reached at stage 36.

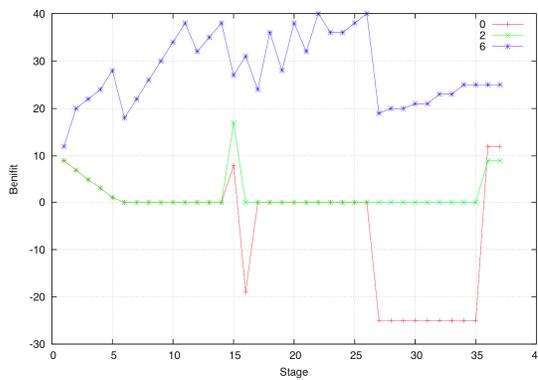 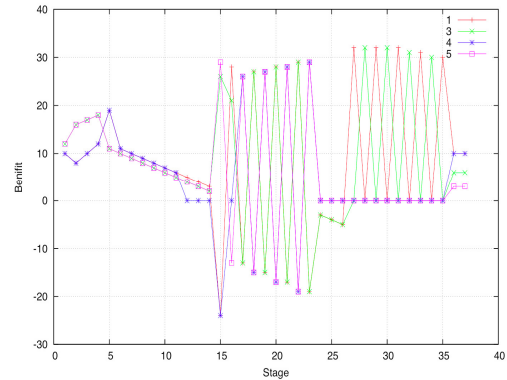

Fig13. The benefits of nodes 0, 2 and 6     Fig14. The benefits of nodes 1, 3, 4 and 5

The figures (Fig15 and Fig16) show the minimum capacity asked by each node at each stage. Each node increases its minimum capacity if it can in order to hope sell routes to its neighbors. Between stages 24 and 25, node 2 decreases its minimum capacity to 0 because the offers that it has received are not satisfactory.

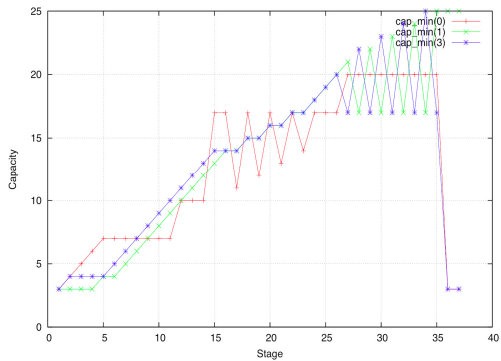

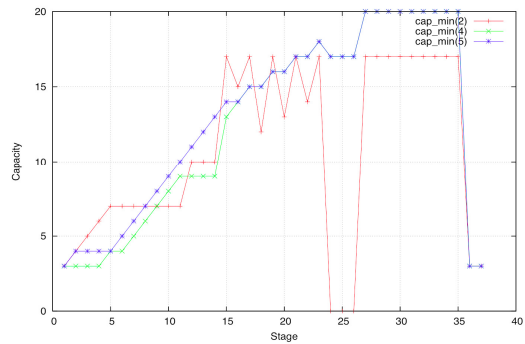

Fig15. The minimum capacity asked by transit providers 0, 1 and 3

Fig16. The minimum capacity asked by transit providers 2, 4 and 5

The figures (Fig17 and Fig18) show at each stage the different capacities of nodes 1 and 3 related to route with node 6 as destination: The minimum capacity asked by this node, its maximum capacity asked, its bought capacity and its used capacity. Recall that the bought capacity by each node is not necessarily used. Our choice is based on both nodes 1 and 3 because they are in competition, and both of nodes want to have the greatest possible number of customers. When the maximum capacity asked by the node (1/or 3) is less than 25, it means that this node has lost against its competitor (3/ or 1) the purchase of more capacity to node 6. Thus this node has asked for this capacity to another of its neighbors that offered a route to destination 6.

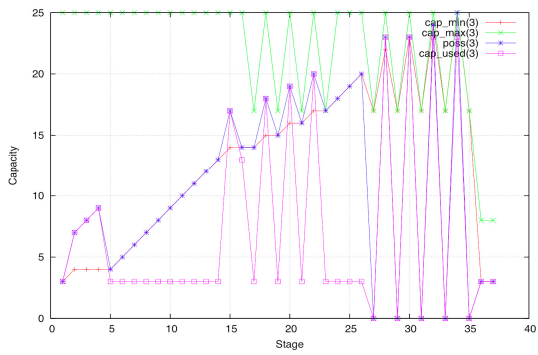

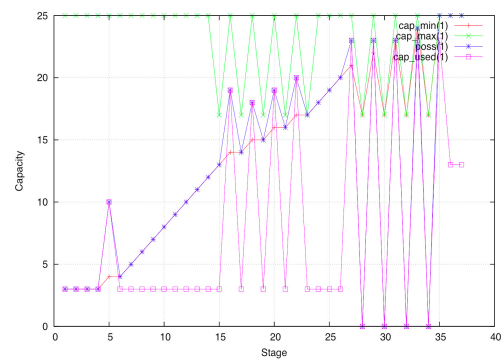

Fig18. The capacities asked, bought and used by node 3

Fig17. The capacities asked, bought and used by node 1

The figure (Fig19) show the different routes obtained from each node in the proposed topology to reach the service desired in the destination node (node 6). These routes are obtained by our simulator at the stable state. In this topology, the satisfaction rate of nodes is (100%), i.e. all nodes are satisfied.

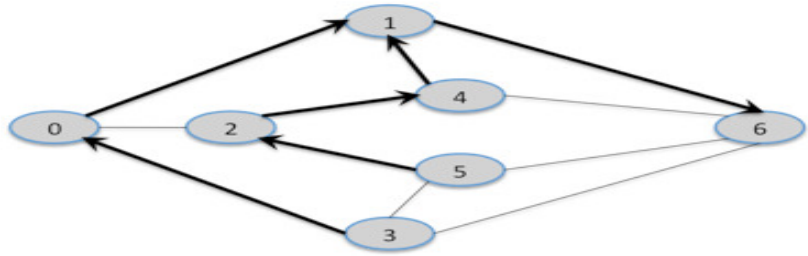

Fig12. The different routes bought by nodes to destination 6 at stable state

## 5. CONCLUSION

We have presented a new model for the QoS and resource management in the interdomain routing framework that we have called stock model. This model is presented using a reverse cascade approach and an iterative process where nodes could learn the right quantities of capacity to buy and stock. The simple simulation results we have presented validate this model. They show that in case of simple network topologies and/or basic strategies, the model converges towards good and satisfying situations. Moreover, since the model converges to positive values of minimum capacities ($cap\_\min$) it doesn't disadvantage providers. But customers are not even more disadvantaged as they can purchase positive stocks of capacity. Therefore, this model really constitutes a stock model that urges on nodes to anticipate the stock of resources that they will be able to resell. However, in cases of more complex network topologies or scenarios (for instance with several destinations), this model would need to be reinforced with more sophisticated features and policies.